\documentclass[reqno]{amsart}

\usepackage{graphicx}
\usepackage{amsfonts}
\usepackage{amsmath}
\usepackage{algorithm} 
\usepackage{subfigure}

\usepackage{hyperref}
\usepackage{xcolor}

\usepackage{tikz}
\usetikzlibrary{calc,shapes,arrows}

\usepackage{algorithm}
\usepackage{algorithmic}

\usepackage{xspace}

\newtheorem{theorem}{Theorem}[section]
\newtheorem{lemma}[theorem]{Lemma}

\newtheorem{definition}[theorem]{Definition}

\numberwithin{equation}{section}

\newcommand{\T}{{\mathbb T}}

\newcommand\blfootnote[1]{%
	\begingroup
	\renewcommand\thefootnote{}\footnote{#1}%
	\addtocounter{footnote}{-1}%
	\endgroup
}

\begin{document}
	
\begin{abstract}
Current cyber-physical systems (CPS) are expected to accomplish complex tasks. To achieve this goal, high performance, but unverified controllers (e.g. deep neural network, black-box controllers from third parties) are applied, which makes it very challenging to keep the overall CPS safe. By sandboxing these controllers, we are not only able to use them but also to enforce safety properties over the controlled physical systems at the same time. However, current available solutions for sandboxing controllers are just applicable to deterministic (a.k.a. non-stochastic) systems, possibly affected by bounded disturbances. In this paper, for the first time we propose a novel solution for sandboxing unverified complex controllers for CPS operating in noisy environments (a.k.a. stochastic CPS). Moreover, we also provide probabilistic guarantees on their safety. Here, the unverified control input is observed at each time instant and checked whether it violates the maximal tolerable probability of reaching the unsafe set. If this probability exceeds a given threshold, the unverified control input will be rejected, and the advisory input provided by the optimal safety controller will be used to maintain the probabilistic safety guarantee. The proposed approach is illustrated empirically and the results indicate that the expected safety probability is guaranteed.
\end{abstract}

\title[Sandboxing Controllers for Stochastic Cyber-Physical Systems]{Sandboxing Controllers for Stochastic Cyber-Physical Systems$^*$}

\author{Bingzhuo Zhong$^{1}$}
\blfootnote{\small *This work was supported in part by the H2020 ERC Starting Grant AutoCPS (grant agreement No 804639) and German Research Foundation (DFG) through the grants ZA 873/1-1 and ZA 873/4-1. Marco Caccamo was supported by an Alexander von Humboldt Professorship endowed by the German Federal Ministry of Education and Research. Any opinions, findings, and conclusions or recommendations expressed in this publication are those of the authors and do not necessarily reflect the views of the Alexander von Humboldt Foundation.}
\author{Majid Zamani$^{3,2}$}
\author{Marco Caccamo$^{1}$}
\address{$^2$Department of Mechanical Engineering, Technical University of Munich, Germany.}
\email{ingzhuo.zhong@tum.de}
\email{mcaccamo@tum.de}
\address{$^2$Department of Computer Science, LMU Munich, Germany.}
\address{$^3$Department of Computer Science, University of Colorado Boulder, USA.}
\email{majid.zamani@colorado.edu}
\maketitle

\section{Introduction}\label{sec:introd}
\begin{figure}
	\includegraphics[width=0.8\textwidth]{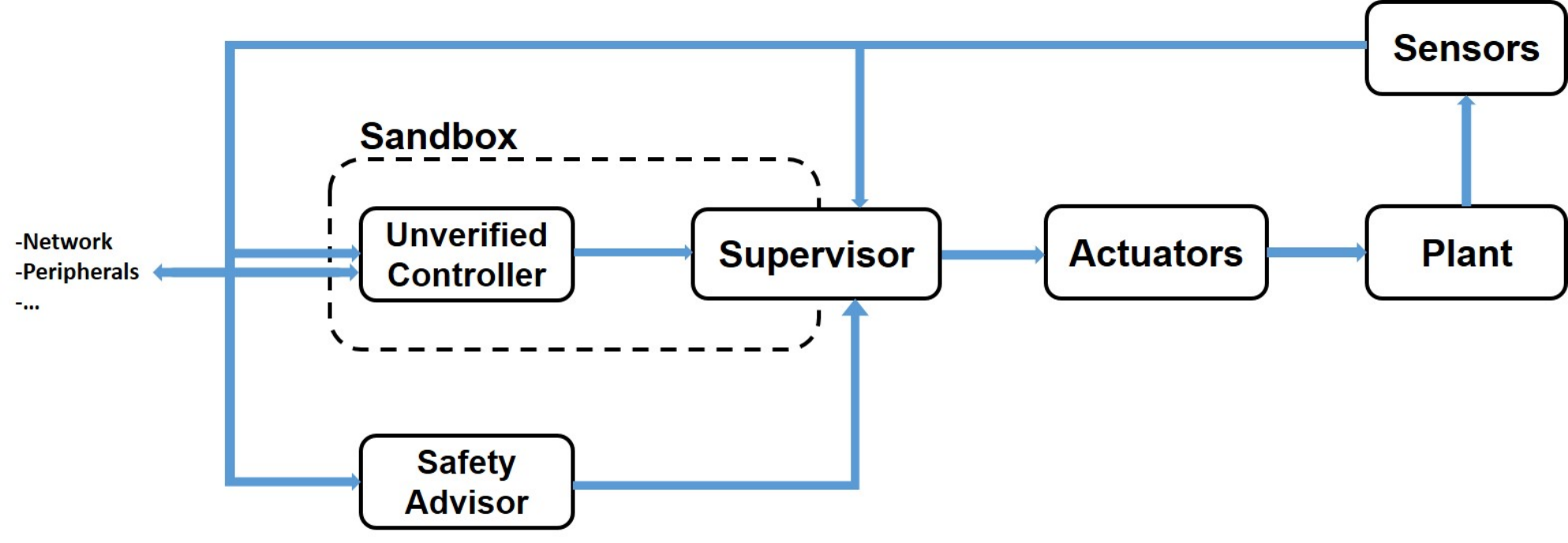}
	\caption{\textbf{Safe}ty Advisor - Super\textbf{visor} (Safe-visor) Architecture for sandboxing unverified \\controller.} \label{fig1:Svisor_arc}
\end{figure}

Cyber-Physical Systems (CPS) are complex systems in which physical components are interacting tightly with cyber ones. These systems are widely used in various kinds of applications, such as automotive, aviation, manufacture plants and so on. Nowadays, these systems are expected to accomplish complex missions. As a result, complex, high performance but unverified controllers (e.g., deep neural network or black-box controllers from third parties) are applied to complete these complex missions, which makes it increasingly challenging to ensure the safety of CPS. To cope with this issue, we exploit the idea of $\mathit{sandbox}$ from the community of computer security, which is a popular security mechanism for cyber systems\cite{Reis2009browser}. In short, it provides a testing environment to isolate the untested and untrusted components from the critical part of a digital controller. The behaviour of the untrusted component is restricted and it can only access the critical part when it follows the rules given by the sandboxing mechanism. 
Hence, we designed a novel architecture that uses a \textbf{Safe}ty Advisor and a Super\textbf{visor} (Safe-visor in short). 
Instead of providing a testing environment and focusing on cyber security, Safe-visor architecture can be used to sandbox any types of unverified controllers in run time regarding the safety of the physical systems. The control inputs of the controller fed to the system are checked and can only be accepted when they are not disobeying the safety rule defined in the sandboxing mechanism. The architecture of safe-visor is illustrated in Figure  \ref{fig1:Svisor_arc}.
In this architecture, the safety of the physical system is characterized by the probability of fulfilling some safety specifications. In general, the Safe-visor specifies verifiable safety rules for the unverified controller to follow so that a specific level of safety probability of the physical system can be ensured. During the execution of the Safe-visor, the Safety Advisor is responsible for providing advisory input for the Supervisor based on the current state of the physical system, which seeks to maximize the safety probability. Meanwhile, the Supervisor checks the input given by the unverified controller according to the safety rule. Input from the unverified controller would only be accepted when it follows the rule; otherwise, the Supervisor would accept the advisory input from the Safety Advisor to maximize the safety probability of the physical system. In the rest of the paper, designing a Supervisor means designing its safety rule for checking the inputs from unverified controllers. It should be noted that inputs given by Safety Advisors only focus on the safety of the system, which should be treated as a fallback in case the unverified controllers are trying to perform some harmful actions. On the other hand, the unverified controller is designed for functionality. i.e. it is expected to realize some tasks which are much more complicated than purely keeping the system safe. By sandboxing the unverified controller, we are able to exploit its advantages for realizing complex tasks while preventing the system from being threatened by its harmful behaviour, if any. 

In this paper, we deal with stochastic CPS modelled as controlled discrete-time Markov process (cdt-MP). We focus on the safety invariance specification, in which the system is expected to stay inside a pre-defined safety set. Here, we formulate the safety invariance specification as a reach-avoid problem in finite time horizon, and design the Safety Advisor based on a finite Markov Decision Process (MDP) constructed from the original cdt-MP.  The inputs given by the unverified controller are checked by the Supervisor at every time instant  based on an estimation of the probability of reaching the unsafe set (i.e., the complement of the pre-defined safety set). 

\subsection*{Related Work}\label{section_related_work}
In \cite{alshiekh2018safe}\cite{humphrey2016synthesis}\cite{bloem2015shield}, a shield is synthesized to correct erroneous output values from those unverified, complexed components in a system so that safety properties can be enforced at run time. This idea is mainly used for systems which can be modelled as automaton, e.g. reactive systems, while our method can be applied to systems with continuous state space and input space. The most relevant work to our proposed method is the one developed based on Simplex architecture \cite{sha2001using,crenshaw2007simplex}, in which the unverified, high-performance controller is sandboxed by an elliptic recovery region associated with a verified, high-assurance controller. Inspired by the idea of sandboxing the unverifiable controllers by using Simplex architecture, many results have been proposed for different kinds of systems and invariance specifications. In the case that bounded uncertainty exists in the system dynamic, L1-Simplex \cite{wang2013l1simplex} is applicable by using L1-adaptive controller \cite{hovakimyan20111} as the high-assurance controller with which the linear model uncertainty in the system dynamic is estimated and compensated. RSimplex \cite{wang2018rsimplex} uses Robust Fault-Tolerant Controller (RFTC) with the similar idea of L1-Simplex, but it is capable of dealing with non-linear model uncertainty. Net-Simplex \cite{yao2013netsimplex} is able to cope with bounded time delay introduced by the network connection in the system. It models the system as a linear parameter-varying system and accordingly designs time-delay-related recovery region. A recent result in \cite{abdi2018preserving} proposes a way to sandbox an unverified controller which may suffer from undetectable cyber attacks by dynamically planning and executing  high-assurance controllers so that the physical system is not endangered. The common point of these results is that Lyapunov-function-based safety invariant sets are used as recovery regions. 

The main difference between the Simplex architecture and our proposed result is that in our proposed solution, only the unverified controller is in charge of accomplishing the task, under the supervision of the Supervisor, rather than designing two parallel controllers (i.e. the high-assurance and high-performance controllers) for the given task and define a verified decision logic to decide which one to be used. The Safety Advisor is not expected to fully control the system and finish the complicated task, but it is only responsible for providing fallback to maximize the safety probability. Then, by properly designing the Supervisor, the unverified controller has more flexibility for functionality, and the safety probability can be guaranteed due to the existence of the fallback solution.

There are some other results which extend the concept of Simplex architecture using reachability analysis to cope with the aforementioned conservativeness. Results in \cite{bak2011sandboxing} provide a backward reachability based method to generate a decision module between the mission controller and the safety one which mainly focuses on the safety of the system. Results in \cite{bak2014real} propose a method in which real-time reachability is integrated into the Lyapunov invariance-based method. This largely increases the feasible region of the mission controller. Another idea to get rid of the conservativeness is to compute the safety invariant purely based on offline reachability analysis, as discussed in \cite{abdi2017application}. It should be noted that these methods are only designed for deterministic (non-stochastic) systems. To the best of our knowledge, our result is the first work with a solution for sandboxing unverified controllers in stochastic settings. 

The rest of the paper is organized as follows: we provide preliminary discussion regarding the notations, models used in this work and formulation of the problem in Section \ref{section_preliminaries}. Then, a scheme to design the Safety Advisor and Supervisor is proposed in Section \ref{sec:Design_of_Svisor}, which will be empirically tested by two case studies in Section \ref{Case_study}. Finally, Section \ref{Conclusion} concludes the paper.

\section{Problem Formulation}\label{section_preliminaries}

\subsection{Preliminaries}\label{subs:preli}
A topological space $S$ is called a Borel space if it is homeomorphic to a Borel subset of a Polish space (i.e., a separable and completely metrizable space). One of the common examples of Borel space are the Euclidean spaces $\mathbb{R}^n$. Any Borel space $S$ is assumed to be endowed with a Borel $\sigma$-algebra denoted by $\mathcal{B}(S)$. A map $f\,:\,X\,\rightarrow\,Y$ is measurable whenever it is Borel measurable. A map $f\,:\,X\,\rightarrow\,Y$ is universally measurable if the inverse image of every Borel set is measurable with respect to every complete probability measure on $X$ that measures all Borel subsets of $X$.\par

For the stochastic kernel, we adopt the notation as in \cite{tkachev2013quantitative}. Given two Borel space $X$ and $Y$, the stochastic kernel on $X$ given $Y$ is the map $P:Y\,\times\,\mathcal{B}(X)\,\rightarrow\,[0,1]$ such that $P(\cdot|y)$ is a probability measure on $X$ for any point $y\in Y$ and $P(B|\cdot)$ is a measurable function on $Y$ for any set $B\,\in \mathcal{B}(X)$.

\subsection{Notations}
We denote by $\mathbb{R}$ the set of real numbers and by $\mathbb{N}$ the set of natural numbers. We denote by $\overline{0,n}:=\,\{0,1,\dots,n\}$ an interval in $\mathbb{N}$  starting from $0$ and ending at $n\in \mathbb{N}$. Set $\mathbb{R}^n$ represents the $n$-dimensional Euclidean space where $n \in \mathbb{N}$. \par

\subsection{Model Description and Problem Formulation}
In this paper, we focus on discrete-time stochastic control systems in the following form:
\begin{equation}
x(t+1)=f(x(t), u(t), w(t)),
\end{equation}
in which $t,t+1$ $\in \overline{0,n}$ are two successive time instants in the time domain $\overline{0,n}$ of the system, where $n\in \mathbb{N}$. Here, $x(t)\in X$ is the state of the system at time $t$, where $X\subseteq\mathbb{R}^n$ is a Borel space as the state space of the system. We denote by $(X, \mathcal{B}(X))$ the measurable space with $\mathcal{B}(X)$ being the Borel sigma-algebra on the state space. We denote by $u(t)\in U$ the input to the system at time $t$, where $U\subseteq\mathbb{R}^m$ is a Borel space as the input space of the system.  We denote by $w(t)$ 
the uncertainty at time instant $t$ where $w:\mathbb{N}\rightarrow\mathbb{R}^d $ is a sequence of independent and identically distributed (i.i.d.) random variables.
Map $f:\, X\,\times\,U\,\times\,\mathbb{R}^d\,\rightarrow\,X$ is a measurable function characterizing the state dynamic of the system. In this paper, we focus on stochastic systems, which can also be formulated as controlled discrete-time Markov Processes (cdt-MP). 
\begin{definition}
	(cdt-MP) \cite{hernandez2012discrete} A controlled discrete-time Markov process is a tuple
	\begin{center}
		$\mathfrak{D}\, =\, (X,\,U,\,\{U(x)\}_{x\in X},\,T_{\mathfrak{D}}),$ 
	\end{center}
	where $X\subseteq\mathbb{R}^n$ is a Borel space representing the state space of the model and $U\subseteq\mathbb{R}^m$ is a Borel space referring to the input space. The set $\{U(x)\}_{x\in X}$ is a family of non-empty measurable subsets of $U$, and $U(x)$ is the set of feasible inputs when system is at state $x$. We denote by $T_{\mathfrak{D}}$ a Borel measurable stochastic kernel $T_{\mathfrak{D}}:\,\mathcal{B}\,\times\,X\,\times\,U\,\rightarrow\,[0,1]$, which assigns to any $x\in X$ and $u\in U(x)$ a probability measure on the Borel space $(X, \mathcal{B}(X))$ and characterizes the state transition of the Markov process. 
\end{definition}  
In the rest of the paper, we focus on systems in which $U(x)=U$, i.e. all inputs are feasible at any state in the evolution of the system. The evolution of the system is described by paths as defined below. 
\begin{definition}
	(Path) A path of a cdt-MP $\mathfrak{D}$ is 
	$$\omega\,=\,(x(0),u(0),x(1),u(1),\ldots,x(t),u(t),\ldots),$$
	where $x(t)\in X$ and $u(t)\in U$, $t\in \overline{0,n}$, $\overline{0,n}\subset \mathbb{N}$ is the time domain of the path. We denote by $\omega_x=\{x(i)\}_{i\in \overline{0,n}}$ and $\omega_u=\{u(i)\}_{i\in \overline{0,n}}$ the subsequences of states and inputs in $\omega$. 
\end{definition}

Given a cdt-MP $\mathfrak{D}$, we are interested in Markov policies to control the system.

\begin{definition}\label{def_mp}
	(Markov policy) For a cdt-MP $\mathfrak{D}\, =\, (X,\,U,\,\{U(x)\}_{x\in X},\,T_{\mathfrak{D}})$, a Markov policy $\mu$ is a sequence $\mu\,=\,(\mu_0,\,\mu_1,\,\mu_2,\ldots)$ of universally measurable map $\mu_t:\,X\,\rightarrow\,U$ at time $t\in \overline{0,n}$, where $\overline{0,n}$ is the time domain of $\mathfrak{D}$.
\end{definition}
With Markov policies, the input at time $t$ is only determined by the state at the same time instant, i.e. $u(t)\,=\,\mu_t(x(t))$. In this paper, we are interested in the safety specification, where the state sequences are expected to stay (with a given probability threshold) inside a safe subset of the state set. We formulate this specification as a reach-avoid problem in finite time horizon. 
\begin{definition} 
	(Reach-avoid problem) Consider a safety set $\mathcal{A}\subset \mathcal{B}(S)$, a bounded Borel set as a safe set, and $\mathcal{A}^c = \mathcal{B}(S)\backslash \mathcal{A}$, as its complement, i.e. an unsafe set. We define the reach-avoid problem under Markov policy $\mu$ over time horizon $\overline{0,N}$ 
	as the following:
	$$p_{s_0}^{\mu}(\mathcal{A}^c) = \mathbb{P}_{s_0}^{\mu}\{ s(k)\,\in\, \mathcal{A}^c | \exists\,k\,\in\, \overline{0,N},\, s(0)\,=\,s_0\},$$
	where $s_0 \in \mathcal{A}$. The minimal probability of reaching the unsafe set is defined as:
	\begin{center}
		$p_{*,s_0}(\mathcal{A}^c) = \inf\{p_{s_0}^{\mu}(\mathcal{A}^c)\, , \mu \in \Pi_{M}^{N}\}$,
	\end{center}
	where $\Pi_{M}^{N}$ is the set of all Markov policies over time horizon $\overline{0,N}$ . A Markov policy $\mu_{*}$ is optimal with respect to an initial state $s_0$ if $p_{s_0}^{\mu_{*}}(\mathcal{A}^c) =p_{*,s_0}(\mathcal{A}^c)$. 
\end{definition}

\section{Design of Safe-visor}\label{sec:Design_of_Svisor}
As discussed in the introduction, designing the Safe-visor for sandboxing unverified controllers consists of designing a Safety Advisor and a Supervisor. The Safety Advisor is designed regarding the safety specification. The Supervisor, on the other hand, is designed for detecting (potential) harmful behaviours of the unverified controller and accordingly deciding the input fed to the system (either the one from the unverified controller or from the Safety Advisor).

Regarding the safety invariance specification, an optimal safety controller can be designed as discussed in \cite{abate2008probabilistic,esmaeil2014formal}, which provides optimal safety policy to guarantee a minimal probability of reaching the unsafe set in a finite time horizon. We use this controller as the Safety Advisor, which is introduced in details in Section \ref{section:safety advisor}.

Since the Safety Advisor only focuses on minimizing the probability of reaching the unsafe set, we need to turn to the unverified controller for functionality. Nevertheless, we still expect a high level of safety for the system. Therefore, we denote by $\rho$ the \textbf{maximal tolerable probability of reaching the unsafe set} that we are able to accept, which quantifies the compromise between functionality and safety. Given $\rho$, the Supervisor can decide whether it should accept inputs from an unverified controller at some time instants by estimating the probability of reaching the unsafe set and compare it with $\rho$. Details for designing a Supervisor working in this way is discussed in Section \ref{sec:supervisor}. It should be noted that the design of unverified controller is not the topic of this paper. The approach proposed here can be applied to any unverified controller as long as the set of all possible inputs provided by the unverified controller is a subset of the input set of the Supervisor. Moreover, we focus on those unverified controllers whose behaviour are unpredictable, i.e. we do not know the exact action of the unverified controller in a given state unless the system actually reaches that state, and the action of unverified controller at the same state may be time dependent. Otherwise, we may be able to verify this controller and sandboxing may not be needed anymore. In the rest of the paper, we denote by $u_{uc}(x,t)$ the input provided by the unverified controller at state $x$ at time instant $t$.

\subsection{Safety Advisor} \label{section:safety advisor}
As mentioned above, we use optimal safety controller with respect to safety invariance specification as the Safety Advisor. To synthesize the optimal safety controller, we define a value function \cite{tkachev2013quantitative}:
\begin{equation}\label{eq:def_vfunction}
V_n^{\pi}(x):=P_x^{\pi}(\diamond^{\leq n}\mathcal{A}^c),
\end{equation}
for $n\in \mathbb{N}$ to denote the probability of reaching the set $\mathcal{A}^c$ in the finite time horizon $\overline{0,n}$ from the initial state $x$, where $\pi\,\in\,\Pi_M$ is a Markov policy and $\diamond^{\leq n}\mathcal{A}^c\,:=\,\{\omega\in \Omega\,:\, \omega_x(k)\in \mathcal{A}^c\,\mbox{for some}\,0 \leq k\leq n\}$, where $\Omega$ is the set of all possible paths within the time horizon $\overline{0,n}$. Since we formulate the safety invariance specification as reach-avoid problem in a finite time horizon, we should minimize the probability mentioned above and, hence, the optimal value function is given by 
\begin{equation}
V_{*,n}(x):=\inf\limits_{\pi \in \Pi}V_n^{\pi}(x),
\end{equation}
initialized with  $V_{*,0}=1_{\mathcal{A}^c}(x)$, where $1_{M}(x)=1$ when $x\in M$ and $1_{M}(x)=0$ otherwise. In fact, this optimal value function can be recursively calculated in the following way (\cite{tkachev2013quantitative}, Corollary 3): 
\begin{equation}
\label{equ:min value function}
V_{*,n+1}(x) = 1_{\mathcal{A}^c}(x)+1_{\mathcal{A}}(x)\inf\limits_{u\in U}\int_{X}^{ }V_{*,n}(y)T_{\mathfrak{D}}(dy|x,u),
\end{equation}
and the optimal policy at time $t=k$ associated to $V_{*,n+1}(x)$ can be obtained as the following 
\begin{equation}
\mu_{*,k}\in \mathop{\arg\inf}_{\mu_k} \int_{X}^{ }(1_{\mathcal{A}^c}(y)+1_{\mathcal{A}}(y)V_{*,n}(y))T_{\mathfrak{D}}(dy|x,\mu_k(x)).
\end{equation}
However, analytical solution of the value function as well as the optimal policy above is very difficult to be obtained in general. Alternatively, we abstract the original cdt-MP $\mathfrak{D}$ and construct a finite Markov Decision Process (MDP) as proposed in \cite{tkachev2013quantitative}, and calculate the solutions based on this finite MDP. First, we use uniform grids to partition the safety region and input set of the cdt-MP. Let $X_p=\bigcup_{i=1}^N \tilde{X}_i$ be a measurable partition of the safety set $A$ and $U_p=\bigcup_{j=1}^M \tilde{U}_j$ a measurable partition of $U$. Let $\tilde{x}_i\in \tilde{X}_i$ for $1\leq i \leq N$ be representative points of $\tilde{X}_i$ and let $\tilde{u}_j\in \tilde{U}_j$ for $1\leq j \leq M$ be representative points of $\tilde{U}_j$. We define the discretization parameter $\delta_x=\max\limits_{\tilde{x}_i,\,\tilde{x}_i'\in X_p}\mathbf{d}_X(\tilde{x}_i,\,\tilde{x}_i')$ for the state set and $\delta_u=\max\limits_{\tilde{u}_j,\,\tilde{u}_j'\in U_p}\mathbf{d}_U(\tilde{u_j},\,\tilde{u}_j')$  for the input set where $\mathbf{d}_X$ and $\mathbf{d}_U$ are the metrics (e.g. Euclidean ones) over sets $X$ and $U$, respectively. Then, the constructed finite MDP is denoted by $\mathfrak{M}\,=\,\{\tilde{X},\,\tilde{U},\,\tilde{T}\}$, in which $\tilde{X} := \{\tilde{x}_i\}_{i=1}^{N}\cup \{\phi\}$, $ \{\tilde{x}_i\}_{i=1}^{N}$ is the set of representative points of $X_p$, $\phi$ is a ``sink" state representing the unsafe set $\mathcal{A}^c$ in the original cdt-MP, and $\tilde{U} := \{\tilde{u}_j\}_{j=1}^{M}$ is the set of representative points of $U_p$. The stochastic kernel $\tilde{T}$ is then a matrix, which can be computed as follows:
$\tilde{T}(\tilde{x}_m|\tilde{x}_i,\tilde{u}_j)=\left\{\begin{aligned} T_{\mathfrak{D}}(\tilde{X}_m|\tilde{x}_i,\tilde{u}_j)\qquad &\text{if $\tilde{x}_i$, $\tilde{x}_m\,\in\,\{\tilde{x}_i\}_{i=1}^{N}$, $\tilde{u}_j\in \tilde{U}$}\\T_{\mathfrak{D}}(\mathcal{A}^c|\tilde{x}_i,\tilde{u}_j)\qquad &\text{if $\tilde{x}_i\,\in\,\{\tilde{x}_i\}_{i=1}^{N}$, $\tilde{x}_m\,\in\,\{\phi\}$, $\tilde{u}_j\in \tilde{U}$}\\1\qquad &\text{if $\tilde{x}_i$, $\tilde{x}_m\,\in\,\{\phi\}$, $\tilde{u}_j\in \tilde{U}$}\\0\qquad &\text{if $\tilde{x}_i\,\in\,\{\phi\}$, $\tilde{x}_m\,\in\,\{\tilde{x}_i\}_{i=1}^{N}$, $\tilde{u}_j\in \tilde{U}$} \end{aligned}\right.$

For the finite MDP $\mathfrak{M}$, we denote by $\tilde{V}_{*,n}(\tilde{x})$ the n-horizon minimal value function for the reach-avoid problem. Similar to equation \eqref{equ:min value function}, we initialize it with $\tilde{V}_{*,0}=1_{\{\phi\}}(\tilde{x})$ and it can be calculated recursively as follows:
\begin{equation}
\label{eq:opt_v_function}
\tilde{V}_{*,n+1}(\tilde{x}) = 1_{\{\phi\}}(\tilde{x})+1_{\{\phi\}^c}(\tilde{x})\min\limits_{\tilde{u}\in \tilde{U}}\sum_{\tilde{y}\in \tilde{X}}\tilde{V}_{*,n}(\tilde{y})\tilde{T}(\tilde{y} |\tilde{x},\tilde{u}),
\end{equation}
and the optimal policy at time $t=k$ associated to $\tilde{V}_{*,n+1}(\tilde{x})$ is given by 
\begin{equation}
\mu_{*,k}(\tilde{x})\in \mathop{\arg\min}_{\tilde{\mu}_k} \sum_{\tilde{y}\in \tilde{X}}(1_{\{\phi\}^c}(\tilde{y})+1_{\{\phi\}}(\tilde{y})\tilde{V}_{*,n}(\tilde{y}))\tilde{T}(d\tilde{y}|\tilde{x},\tilde{\mu}_k(\tilde{x})).
\end{equation}

In principle, the optimal policy can be obtained for arbitrary long time horizon, but $\tilde{V}_{*,n}(\tilde{x})$ will keep decreasing, i.e. the probability of avoiding the unsafe set is decreasing, when $n$ increases. Therefore, the time horizon of the optimal policy cannot be arbitrarily long, but it is tunable up to some degrees by setting the maximal tolerable value of the value function, i.e. the smaller (bigger) the maximal tolerable value of the value function is, the shorter (longer) the time horizon for the optimal safety policy is. This value should not be bigger than the maximal tolerable probability of reaching the unsafe set, i.e. $\rho$, as defined in the beginning of Section \ref{sec:Design_of_Svisor}, so that $\rho$ can be guaranteed at least by accepting advisory input from the Safety Advisor. Therefore, in our implementation, the time horizon $\overline{0,H}$ of the Safety Advisor is determined in a way such that $\forall \tilde{x}\in\tilde{X}\backslash\{\phi\}, \tilde{V}_{*,H}(\tilde{x})\leq \rho$ and $\exists \tilde{x}\in\tilde{X}\backslash\{\phi\}, \tilde{V}_{*,H+1}(\tilde{x}) > \rho$.

\subsection{Supervisor}\label{sec:supervisor}
As previously mentioned, the Supervisor is required to estimate the probability of reaching the unsafe set, in case it accepts inputs from the unverified controller. Since the safety guarantee given by the Safety Advisor is calculated based on the abstraction of the original stochastic system, i.e. the finite MDP, for consistency of guarantee regarding safety probability, we use the same finite MDP to design the Supervisor. 

As discussed in the previous section, the probability of reaching the unsafe set of the finite MDP is quantified by value function $\tilde{V}_H^{\mu}(s_0)$ according to equation \eqref{eq:def_vfunction}. When the initial state $s_0$ and the time horizon $\overline{0,H}$ are fixed, the value function is varied by different $\mu$. Meanwhile, compared with purely using the optimal safety policy $\mu_{*}$, sandboxing the unverified controller and accepting it at some states at some time instants intrinsically result in a new Markov policy for controlling the system, according to the architecture of Safe-visor. Therefore, to ensure that the probability of reaching the unsafe set is lower than the predefined $\rho$ in a given time horizon $\overline{0,H}$, the Supervisor should be designed in a way such that the following inequality holds: 
\begin{equation}
\label{req_for_sp}
\tilde{V}_H^{\mu'}(s_0)\le \rho,
\end{equation}
where $\mu'$ is the Markov policy used to control the system, when the unverified controller is accepted at some states at some time instants by the Supervisor.
In Section \ref{section:safety advisor}, the optimal safety policy is obtained by selecting a Markov policy minimizing the value function of each state at each time instant. In other way, when the Markov policy is fixed, we can calculate the value function in the way illustrated in the next theorem.

\begin{theorem}\label{theorem_vfunction}
	Given a Markov policy $\mu\,=\,(\mu_0, \mu_1,\ldots,\mu_{H-1})$ in a finite time horizon $\overline{0,H}$, the value function $\tilde{V}_{n}(\tilde{x})$ can be recursively calculated in the following way:
	\begin{equation}
	\tilde{V}_{n+1}(\tilde{x}) = 1_{\{\phi\}}(\tilde{x})+1_{\{\phi\}^c}(\tilde{x})\sum_{\tilde{y}\in \tilde{X}}\tilde{V}_{n}(\tilde{y})\tilde{T}(\tilde{y} |\tilde{x},\mu_{H-n-1}(\tilde{x})),
	\end{equation}
	where $\tilde{x} \in \tilde{X}$ and $\tilde{V}_{0}(x)=\tilde{V}_{*,0}(x)$. 
\end{theorem}
Theorem \ref{theorem_vfunction} can be proved similar to the proof of Lemma 1 in \cite{abate2008probabilistic}, since Lemma 1 in \cite{abate2008probabilistic} can be treated as a general case for Theorem \ref{theorem_vfunction}. With having Theorem \ref{theorem_vfunction}, the remaining question is how to determine $\mu'$ at run time. Let $\mu'=\,(\mu_{0}',\mu_{2}',\ldots,\mu_{H-1}')$. When the Supervisor is being executed, at every time instant $k \in \overline{0,H-2}$, $\mu'_t$ are unknown for all $t$ where $k<t\leq H-1$ (i.e.,  the Markov policy used to control the system in the future time is unknown). To guarantee the  safety threshold specified by $\rho$, at every time instant $k\in\overline{0,H-2}$, input from the unverified controller can only be accepted, when inequality \eqref{req_for_sp} is at least fulfilled in the case that the Supervisor only accepts the advisory input from the safety advisor afterwards. This requirement is formally defined in Definition \ref{PR_for_UC}.

\begin{definition}\label{PR_for_UC}
	Given current time instant $k$, where $0 \leq k \leq H-2$, $\omega$ is the path up to $k$ and $\rho$ is the maximal tolerable probability of reaching the unsafe set, the input $u_{uc}(\omega_x(k), k)$ from the unverified controller can only be accepted, if there exists a Markov policy $\mu =\{\mu_0,\,\mu_1,\,\mu_2, \ldots, \mu_{H-1}\}\in M$ such that $\tilde{V}_H^{\mu}(\omega_x(0))\le \rho$, where $M$ denotes the set of all Markov policies, $\mu_k(\omega_x(k))=u_{uc}(\omega_x(k), k)$, and for all $t$ where $k<t\leq H-1$, $\mu_t = \mu_{*,t}$.
\end{definition}

In general, it is difficult to calculate the exact value of $\tilde{V}_H^{\mu'}(s_0)$ at run time due to the lack of adequate information from the past. At each time instant $k$ during the execution, where $k\in \overline{0,H-2}$, the only available information for the Supervisor is the path $\omega$ of the system up to $k$. In other words, the Supervisor does not have complete information about $\mu'_t$ for all $t\in \overline{0,k}$, since $\mu'_t(\tilde{x})$ is unknown when $\tilde{x} \in \tilde{X}\backslash\{\omega_x(t)\}$. 
To cope with this difficulty, we propose a novel Supervisor, namely \textit{History-based Supervisor}, as defined in Definition \ref{def:History-based Supervisor}, which is able to check the feasibility of the input provided by the unverified controller only based on the history information during the execution (i.e., path $\omega$ of the system up to the current time instant $k$ during the execution).  

\begin{definition}\label{def:History-based Supervisor}
	(History-based Supervisor) For all $k\in \overline{0,H-1}$ \footnote{No input needed to be provided at $t=H$ since it is the end of the execution.}, given the history of path $\omega$ up to $k$, the input $u_{uc}(\omega_x(k), k)$ from the unverified controller can only be accepted, when quantity
	\begin{small}
		\begin{equation*}
		\prod_{t=1}^{k}\sum_{\tilde{x}\in \tilde{X}\backslash\{\phi\}}\tilde{T}(\tilde{x}|\omega_x(t-1), \omega_u(t-1))\left(1-\sum_{\tilde{x}\in \tilde{X}}\tilde{V}_{*,H-k-1}(\tilde{x})\tilde{T}(\tilde{x}|\omega_x(k),u_{uc}(\omega_x(k), k))\right)
		\end{equation*}
	\end{small}is not smaller than $1-\rho$, where $\rho$ is the maximal tolerable probability of reaching the unsafe set.
\end{definition}

By using History-based Supervisor in Safe-visor architecture, it can be guaranteed that $\mu'$ fulfils inequality \eqref{req_for_sp}, as illustrated in the next theorem.

\begin{theorem} \label{theorem:guarantee for hb supervisor}
	Given a finite MDP and the unsafe set ${A}^c$, by using History-based Supervisor at $t$ for all $t\in \overline{0,H-1}$ in Safe-visor architecture, we have 
	\begin{equation}
	p^{\mu'}_{s_0}(\diamond^{\leq H}\mathcal{A}^c)\le \rho, \nonumber
	\end{equation}
	where $\mu'$ is the Markov policy used to control the system when History-based Supervisor is applied. 
\end{theorem}
Proof of Theorem \ref{theorem:guarantee for hb supervisor} is provided in the appendix.

Note that $\tilde{V}_{*,n}$ and $\tilde{T}$ are calculated offline when synthesizing the Safety Advisor. Hence, the Supervisor defined in Definition \ref{def:History-based Supervisor} can be readily used in real-time, since the required computation can be efficiently performed. Concretely, at every time instant $k$ during the execution:

\begin{enumerate}
	\item The number of operations required for computing $\prod_{t=1}^{k}\sum_{\tilde{x}\in \tilde{X}\backslash\{\phi\}}\tilde{T}(\tilde{x}|\omega_x(t-1), \omega_u(t-1))$ is constant, since 
	\begin{small}
		\begin{equation*}
		\begin{split}
		&\prod_{t=1}^{k}\sum_{\tilde{x}\in \tilde{X}\backslash\{\phi\}}\tilde{T}(\tilde{x}|\omega_x(t-1), \omega_u(t-1)) \\
		=&\sum_{\tilde{x}\in \tilde{X}\backslash\{\phi\}}\tilde{T}(\tilde{x}|\omega_x(k-1), \omega_u(k-1)) \times \prod_{t=1}^{k-1}\sum_{\tilde{x}\in \tilde{X}\backslash\{\phi\}}\tilde{T}(\tilde{x}|\omega_x(t-1), \omega_u(t-1))\\
		=&(1-\tilde{T}(\phi|\omega_x(k-1),\omega_u(k-1)) \times \prod_{t=1}^{k-1}\sum_{\tilde{x}\in \tilde{X}\backslash\{\phi\}}\tilde{T}(\tilde{x}|\omega_x(t-1), \omega_u(t-1))
		\end{split}
		\end{equation*}
	\end{small}while $\prod_{t=1}^{k-1}\sum_{\tilde{x}\in \tilde{X}\backslash\{\phi\}}\tilde{T}(\tilde{x}|\omega_x(t-1), \omega_u(t-1))$ has already been computed at the previous time instant (i.e., $k-1$), and $\tilde{T}(\phi|\omega_x(k-1),\omega_u(k-1))$ can be directly obtained from $\tilde{T}$.
	
	\item The number of operations required for computing 
	\begin{small}
		\begin{equation*}
		1-\sum_{\tilde{x}\in \tilde{X}}\tilde{V}_{*,H-k-1}(\tilde{x})\tilde{T}(\tilde{x}|\omega_x(k),u_{uc}(k))
		\end{equation*}
	\end{small}
	is proportional to the number of states of the finite MDP, since $\tilde{V}_{*,H-k-1}(\tilde{x})$ and $\tilde{T}(\tilde{x}|\omega_x(k),u_{uc}(k))$ can directly be obtained in $\tilde{V}_{*,n}$ and $\tilde{T}$.  
\end{enumerate}
The real time applicability of the proposed Supervisor is shown in the experiments in Section \ref{Case_study}.

\section{CASE STUDY}\label{Case_study}
In this section, we apply our approach to two case studies. The first case study is a temperature control problem and the second one is a traffic control problem. We simulate each test case $1.0 \times 10^6$ times and analyze accordingly the percentage of paths staying in the safety set in the given time horizon. For comparison, we simulate these test cases by 1) only using the unverified controller and 2) only using the proposed safety advisor. Moreover, we compute the average execution time for our Supervisor in both cases to show feasibility of running it in real-time. The simulation in this section is performed in MATLAB 2018b, on a computer equipped with Intel(R) Xeon(R) E-2186G CPU (3.8 GHz) and 32 GB of RAM running Window 10.

\subsection{Temperature Control Problem}
In the temperature control problem, a room is equipped with a heater being controlled and the temperature of the room is required to be kept between 19 and 21$^{\circ}$C.  The temperature of the room can be modelled as the following, which is adapted from \cite{lavaei2018dissipativity}:
\begin{equation}
x(k+1) = (1-\beta-\gamma u(k))x(k)+\gamma T_h u(k) + \beta T_e + \omega(k)
\end{equation}
where $x(k)$ denotes the temperature at time $t=k$. Input $u(k)$ takes any real value between 0 to 0.6. Parameter $\beta$ is conduction factor between the external environment and the room, $\gamma$ is conduction factor between the heater and the room, $T_e$ is the temperature of the external environment and $T_h$ is the temperature of the heater. We denote by $\omega$ a Gaussian white noise.  In this section, we set $\beta = 0.022$, $\gamma = 0.05$, $T_e\,=\,-1$, $T_h\,=\,50$, the mean of $\omega$ is 0 and variance is 0.04. The sampling time interval in this example is 9 minutes.

Now, we synthesize the Safety Advisor as discussed in Section \ref{section:safety advisor}. We use the discretization parameter $\delta_x= 1.0\times 10^{-3} $ and $\delta_u = 2.4\times 10^{-2}$ to discretize the safety set (resulting in 2000 discrete states) and the input set (resulting in 25 discrete inputs) to construct a finite MDP. We set $\rho$ as 1\% and obtain a controller for time horizon $\overline{0,40}$ (6 hours). We set the initial state at 19.01$^{\circ}$C. The unverified controller tries to keep the heater idle at all time, i.e. $U_{uc}(t) \equiv 0$ for all $t\in\overline{0,39}$. This is an unacceptable input which cools down the room to an unacceptable low level. 
For the given $\rho$, it is expected that at least 99\% of the paths stay inside the safety set in the given time horizon. The result of the simulation is shown in Table \ref{tab1} and Figure \ref{fig1:TC1}. The temperature keeps decreasing and all paths go outside of the safety set, when the system is fully controlled by the unverified controller. Meanwhile, more than 99\% of the paths stay within the safety set when our proposed method is applied.

\subsection{Traffic Control Problem}
\begin{figure}
	\centering
	\setlength{\abovecaptionskip}{0.cm}
	\includegraphics[width=4cm]{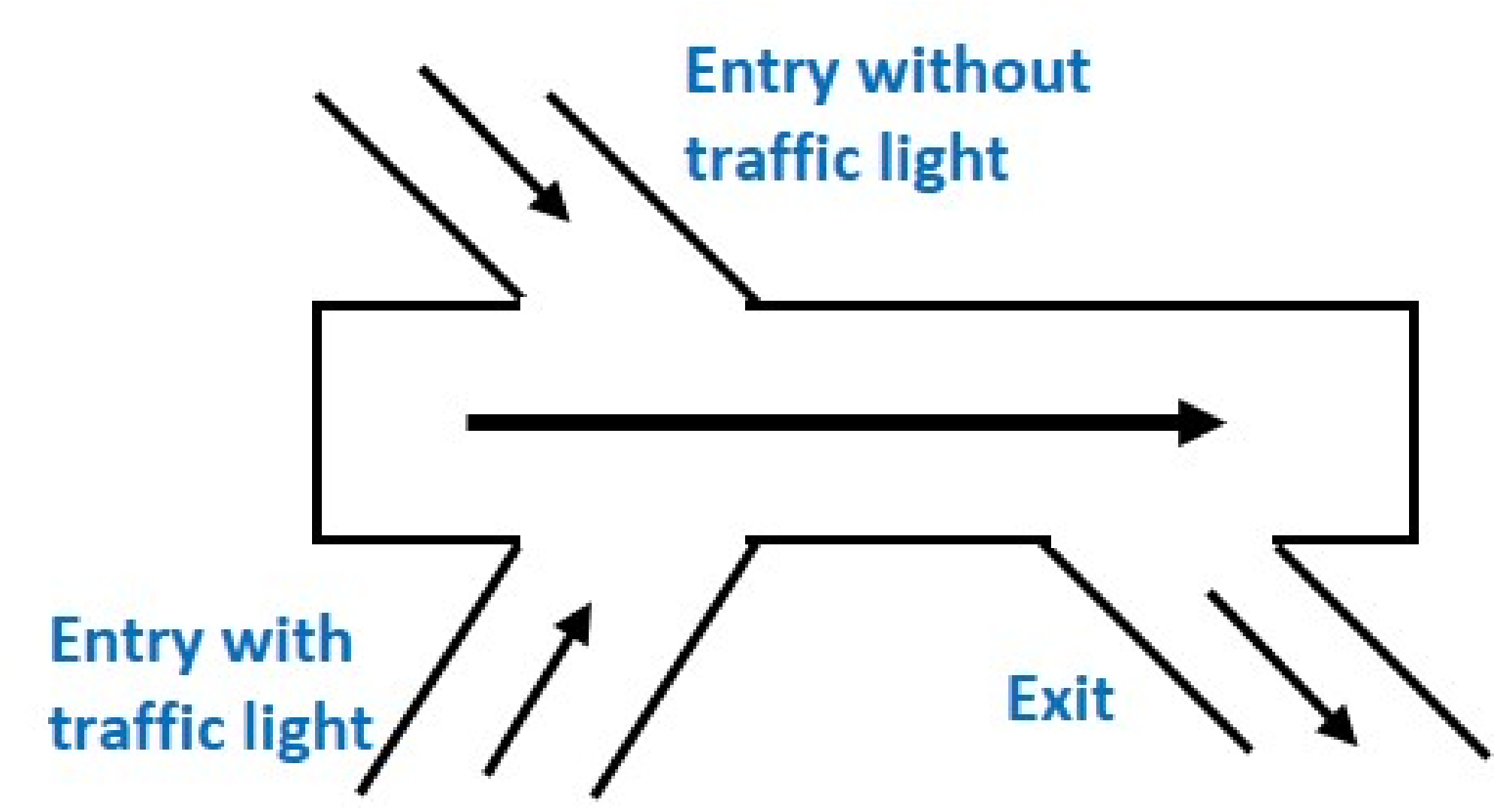}
	\caption{Traffic Control Problem} \label{fig1:Traffic_control_problem}
\end{figure}

\begin{figure}[!]
	\centering	
	\setlength{\abovecaptionskip}{0.cm}
	\includegraphics[width=10cm]{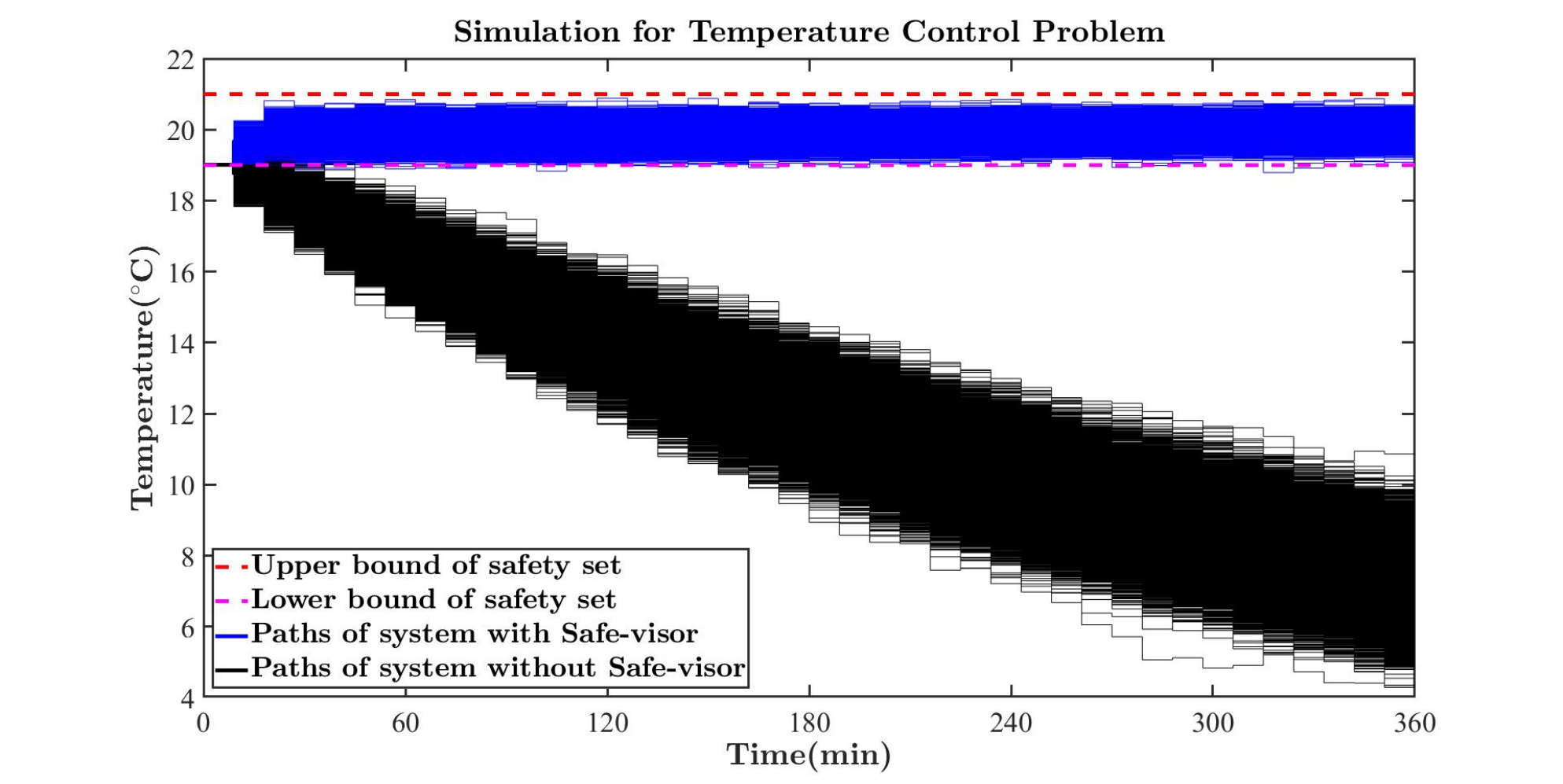}
	\caption{Comparison between paths of system with and without Safe-visor (Temperature Control Problem). } \label{fig1:TC1}
\end{figure}

In the traffic control problem, we focus on a road traffic control containing a cell with 2 entries and 1 exit, as illustrated in Figure \ref{fig1:Traffic_control_problem}. 
One of the entry is controlled by a traffic light. The dynamic of the system can be modelled as the following, which is adapted from \cite{lavaei2019compositional}:
\begin{equation}
x(k+1) = (1-\frac{\tau v}{l}-q)x(k)+e_1u(k)+ \sigma(k)+e_2,
\end{equation}
where $x(k)$ denotes the density of traffic at time $k$, $u(k)\in\{0,1\}$ is the input to the system (1 means the green light is on while 0 means the red light is on). Parameter $v$ is the flow speed of the vehicle on the road, $l$ is the length of the cell, $\sigma$ is a white Gaussian noise, and $\tau$ denotes the sampling time interval of the system. In one sampling interval, $e_1$ is the number of cars that pass the entry controlled by the traffic light, $e_2$ refers to the number of cars that pass the entry without traffic light, and $q$ is the percentage of cars which leave the cell through the exit. In the simulation, we set $l = 500$[m], $v = 25$[m/s], $\tau = 6$s, $e_1=3$, $e_2=6$, $q = 10\%$, the mean of $\sigma$ is $0$ and variance is $2$. In this case study, it is desired that the density of traffic is lower than $20$.

\begin{figure}[!]
	\centering	
	\setlength{\abovecaptionskip}{0.cm}
	\includegraphics[width=10cm]{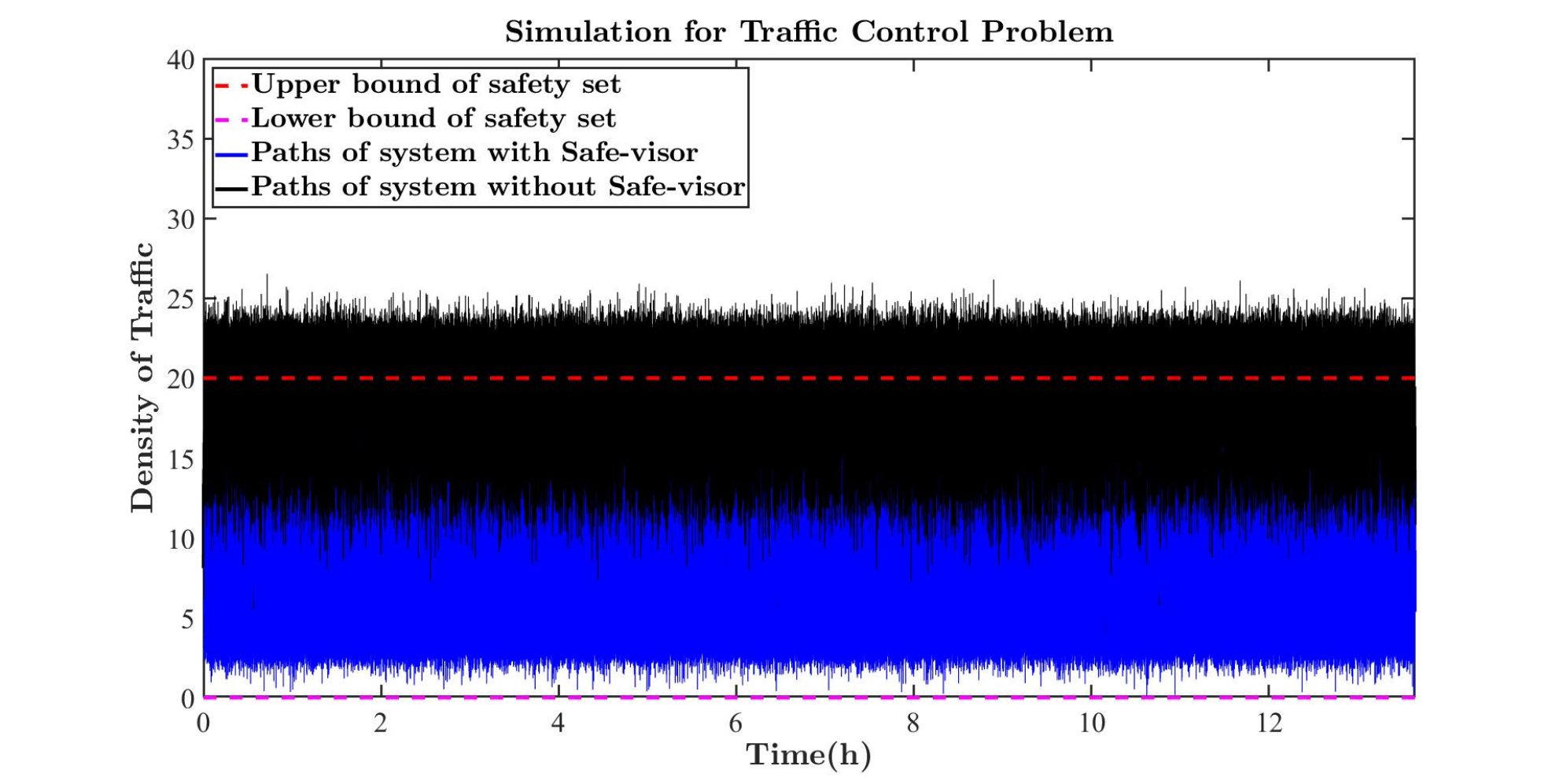}
	\caption{Comparison between paths of system with and without Safe-visor (Traffic Control Problem)} \label{fig1:TC2}
\end{figure}

Now, we synthesize the Safety Advisor as discussed in Section \ref{section:safety advisor}. We use the discretization parameter $\delta_x= 1.0\times 10^{-3} $ to discretize the safety set (resulting in 20000 discrete states) to construct a finite MDP. Note that the input set is already finite. We set $\rho$ as 0.05\% and obtain a controller for the time horizon $\overline{0,8186}$ (13.64 hours). For the simulation, we set the initial state at $x=9$, and choose the unverified controller as the following: $u_{uc}(t)=0$ when $t\in \overline{0,8186}$ is an odd number and $u_{uc}(t)=1$ otherwise. For the given $\rho$, it is expected that at least 99.95\% of the paths stay inside the safety set in the given time horizon. The result of the simulations is shown in Table \ref{tab1} and Figure \ref{fig1:TC2}. All paths go outside of the safety set, when the system is fully controlled by the unverified controller. Meanwhile, more than 99.95\% of paths stay within the safety set when our proposed method is applied.

\begin{table}
	\caption{Result of simulation for both case studies. }\label{tab1}
	\begin{tabular}{|p{8.5cm}|p{2cm}|p{1.5cm}|}
		\hline
		&  Temperature Control & Traffic Control \\
		\hline
		Percentage of paths in the safety set (with Safe-visor)& 99.02\% & 99.958\%\\
		\hline
		Average acceptance rate of the unverified controller & 19.12\% & 8.5114\%\\
		\hline
		Percentage of paths in the safety set (without Safe-visor)&  0\% & 0\%\\
		\hline
		Percentage of paths in the safety set\par (when system is fully controlled by the Safety Advisor) & 99.18\% & 99.989\%\\
		\hline
		Average execution time for the History-based Supervisor & 33.42 $\mu s$ & 31.83 $\mu s$\\
		\hline
	\end{tabular}
\end{table}

According to the empirical result, by sandboxing the unverified controller with Safe-visor architecture, the probabilistic guarantees are respected while some of the inputs from the unverified controller are still accepted for functionality. The average execution time for the History-based Supervisor shows its good real-time applicability, which makes it practical to be applied in real time.

\section{CONCLUSION AND FUTURE WORK}\label{Conclusion}
In this paper, we developed a new framework for sandboxing unverified controllers for stochastic cyber-physical systems regarding safety invariance specification. In comparison with the Simplex architecture, our framework is applicable to stochastic systems, and provides more flexibility for the unverified controllers to accomplish complex tasks.
According to the empirical results for two case studies, the pre-proposed safety probability is guaranteed by using our method. In the future, we would like to extend this method to 1) systems modelled by  Partially Observable Markov Decision Processes\cite{monahan1982state} 2) more general safety specifications, e.g. those expressed as co-safe linear temporal logic formulae \cite{kupferman2001model}.

\section{ACKNOWLEDGEMENTS}
The authors would like to thank Abolfazl Lavaei for the discussions on synthesizing optimal safety controllers for stochastic systems. 

\section{Appendix: Proof of Theorem \ref{theorem:guarantee for hb supervisor}}\label{proof}
The proof of theorem \ref{theorem:guarantee for hb supervisor} is done with the help of the following lemma.
\begin{lemma}
	Given a finite MDP $\mathfrak{M}\,=\,\{\tilde{X},\,\tilde{U},\,\tilde{T}\}$ and a Markov policy $\mu\,=\,(\mu_0, \mu_1,\ldots,\mu_{H-1})$ in a finite time horizon $\overline{0,H}$,  we have
	\begin{small}
		\begin{equation}
		1-\tilde{V}_{n+1}(\tilde{x}) = \sum_{\tilde{y}\in \tilde{X}\backslash\{\phi\}}(1-\tilde{V}_{n}(\tilde{y}))\tilde{T}(\tilde{y}|\tilde{x},\mu_{H-n-1}(\tilde{x}))\nonumber
		\end{equation}
	\end{small}
	where $\tilde{V}_{n}(\tilde{x})$ is the value function for the reach-avoid problem and $\tilde{x}\in \tilde{X}$. 
\end{lemma}

The proof can be readily derived based on Theorem \ref{theorem_vfunction} and the definition of $\tilde{T}$. 
Let $\mu'$ be the Markov policy used to control the system when the unverified controller is accepted at some states at some time instants. Here, we use $\tilde{X}_s$ to represent $\tilde{X}\backslash\{\phi\}$. Let's define:
\begin{equation*}
f(\tilde{x}(k),\mu'_{k}(\tilde{x}(k))) = 1-\sum_{\tilde{x}(k+1)\in \tilde{X}_s}\tilde{V}_{*,H-k-1}(\tilde{x}(k+1))\tilde{T}(\tilde{x}(k+1)|\tilde{x}(k),\mu'_{k}(\tilde{x}(k))),
\end{equation*}
and
\begin{equation*}
g(\tilde{x}(k-1),\mu'_{k-1}(\tilde{x}(k-1)))=\tilde{T}(\tilde{x}(k)|\tilde{x}(k-1),\mu'_{k-1}(\tilde{x}(k-1))).
\end{equation*}
Given initial state $s_0\in \tilde{X}_s$, at each time instant $t=k$ where $k\in \overline{0,H-1}$, we have 
\begin{small}
	\begin{equation*}
	\begin{split}
	&1-\tilde{V}^{\mu'}_{H}(s_0)\\
	=&\sum_{\tilde{x}(1)\in \tilde{X}_s}\left(\sum_{\tilde{x}(2)\in \tilde{X}_s}\left(\ldots\left(\sum_{\tilde{x}(k)\in \tilde{X}_s}f(\tilde{x}(k),\mu'_{k}(\tilde{x}(k)))g(\tilde{x}(k-1),\mu'_{k-1}(\tilde{x}(k-1)))\right)\right.\right.\\
	&\left.\left. \ldots \right)	g(\tilde{x}(1),\mu'_{1}(\tilde{x}(1)))\right)g(s_0,\mu'_{0}(s_0))\\
	\end{split}
	\end{equation*}
\end{small}
\begin{small}
	\begin{equation*}
	\begin{split}
	\geq&\sum_{\tilde{x}(1)\in \tilde{X}_s}\left(\sum_{\tilde{x}(2)\in \tilde{X}_s}\left(\ldots\left(f(\underline{\tilde{x}(k)},\underline{\mu'_{k}(\tilde{x}(k))})\sum_{\tilde{x}(k)\in \tilde{X}_s}g(\tilde{x}(k-1),\mu'_{k-1}(\tilde{x}(k-1)))\right)\right.\right.\\
	&\left.\left. \ldots \right)g(\tilde{x}(1),\mu'_{1}(\tilde{x}(1)))\right)g(s_0,\mu'_{0}(s_0))\\
	\geq&\sum_{\tilde{x}(1)\in \tilde{X}_s}\left(\sum_{\tilde{x}(2)\in \tilde{X}_s}\left(\ldots\left(\left(\sum_{\tilde{x}(k)\in \tilde{X}_s}g(\underline{\tilde{x}(k-1)},\underline{\mu'_{k-1}(\tilde{x}(k-1))}))\right)f(\underline{\tilde{x}(k)},\underline{\mu'_{k}(\tilde{x}(k))})\right.\right.\right.\\
	&\left.\left.\left. \sum_{\tilde{x}(k-1)\in \tilde{X}_s}g(\tilde{x}(k-2),\mu'_{k-2}(\tilde{x}(k-2)))\right) \ldots \right)			g(\tilde{x}(1),\mu'_{1}(\tilde{x}(1)))\right)g(s_0,\mu'_{0}(s_0))\\
	\geq&\sum_{\tilde{x}(1)\in \tilde{X}_s}\left(\sum_{\tilde{x}(2)\in \tilde{X}_s}\left(\ldots\left(\left(\sum_{\tilde{x}(k-1)\in \tilde{X}_s}g(\underline{\tilde{x}(k-2)},\underline{\mu'_{k-2}(\tilde{x}(k-2))})\right)\right.\right.\right.\\
	&\left(\sum_{\tilde{x}(k)\in \tilde{X}_s}g(\underline{\tilde{x}(k-1)},\underline{\mu'_{k-1}(\tilde{x}(k-1))})\right)f(\underline{\tilde{x}(k)},\underline{\mu'_{k}(\tilde{x}(k))})\\	
	&\left.\left.\left. \sum_{\tilde{x}(k-2)\in \tilde{X}_s}g(\tilde{x}(k-3),\mu'_{k-3}(\tilde{x}(k-3)))\right) \ldots \right)			g(\tilde{x}(1),\mu'_{1}(\tilde{x}(1)))\right)g(s_0,\mu'_{0}(s_0))\\	
	&\ldots\\
	\geq& \prod_{t=1}^{k}\sum_{\tilde{x}(t)\in \tilde{X}_s}g(\underline{\tilde{x}(t-1)}, \underline{\mu_{t-1}(\tilde{x}(t-1))}) (f(\underline{\tilde{x}(k)},\underline{\mu'_{k}(\tilde{x}(k))})
	\end{split}
	\end{equation*}
\end{small}
where 
\begin{small}
	\begin{equation*}
	(\underline{\tilde{x}(t-1)},\,\underline{\mu_{t-1}(\tilde{x}(t-1))}) = \mathop{\arg\min}_{\substack{\tilde{x}(t-1)\in \tilde{X}_s\\ \mu_{t-1}(\tilde{x}(t-1))}}\sum_{\tilde{x}(t)\in \tilde{X}_s}g(\tilde{x}(t-1), \mu_{t-1}(\tilde{x}(t-1)))
	\end{equation*}
\end{small}
for all $t\in \overline{0,k}$, and 
\begin{small}
	\begin{equation*}
	(\underline{\tilde{x}(k)},\,\underline{\mu_{k}(\tilde{x}(k))}) = \mathop{\arg\min}_{\substack{\tilde{x}(k)\in \tilde{X}_s\\ \mu_{k}(\tilde{x}(k))}}f(\tilde{x}(k),\mu'_{k}(\tilde{x}(k))).
	\end{equation*} 
\end{small}

Noted that $\omega = (\underline{\tilde{x}(0)}, \underline{\mu_0(\tilde{x}(0))},\underline{\tilde{x}(1)}, \underline{\mu_1(\tilde{x}(1))}\ldots\,\underline{\tilde{x}(k)})$ is one of the paths up to time instant $k$ which can be generated by the system controlled by the Markov policy $\mu'$, and the History-based Supervisor ensures that for all paths $\omega$ up to arbitrary time instant $k\in\overline{0,H}$, 
\begin{equation}
\prod_{t=1}^{k}\sum_{\tilde{x}\in \tilde{X}_s}g(\omega_x(t-1), \omega_u(t-1))\left(f(\omega_x(k),u_{uc}(\omega_x(k),k))\right)\geq 1-\rho. \nonumber
\end{equation}
Note that we have $1-\tilde{V}^{\mu'}_{H}(s_0)\geq 1-\rho$, i.e. $p^{\mu'}_{s_0}(\diamond^{\leq H}\mathcal{A}^c) =\tilde{V}^{\mu'}_{H}(s_0) \leq \rho$.

\bibliographystyle{alpha}
\bibliography{myref}

\end{document}